\documentclass[11pt]{article}
\usepackage{amsfonts}
\usepackage{amssymb,amsmath,mathrsfs}
\usepackage{graphicx}
\usepackage{subfigure}
\begin{document}
\title{\bf{Soliton solutions of an integrable nonlocal modified Korteweg-de Vries equation through inverse scattering transform}}
\author{Jia-Liang Ji and Zuo-Nong Zhu
\footnote{Corresponding author. Email: znzhu@sjtu.edu.cn}
\\
Department of Mathematics, Shanghai Jiao Tong University,\\ 800 Dongchuan Road, Shanghai, 200240, P. R. China\\
}
\date{ }
\maketitle
\begin{abstract}
  It is well known that the nonlinear Schr\"odinger (NLS) equation is a very important integrable equation. Ablowitz and Musslimani introduced and investigated an integrable nonlocal NLS equation through inverse scattering transform. Very recently, we proposed an integrable nonlocal modified Korteweg-de Vries equation (mKdV) which can also be found in a paper of Ablowitz and Musslimani. We have constructed the Darboux transformation and soliton solutions for the nonlocal mKdV equation. In this paper, we will investigate further the nonlocal mKdV equation. We will give its exact solutions including soliton and breather through inverse scattering transformation. These solutions have some new properties, which are different from the ones of the mKdV equation.\\
\end{abstract}
\section{Introduction}
As is well known, the nonlinear Schr\"odinger (NLS) equation
\begin{equation}\label{a1}
iq_t(x,t)=q_{xx}(x,t)\pm2|q(x,t)|^2q(x,t)
\end{equation}
has been investigated deeply since the important work of Zakharov and Shabat \cite{i9}. In physical fields, the NLS equation can characterize plenty of models in varies aspects, such as nonlinear optics \cite{i10}, plasma physics \cite{i12}, deep water waves \cite{i11} and in purely mathematics like motion of curves in differential geometry \cite{i13}.
In fact, the NLS equation can be derived from the theory of deep water wave, and also from the Maxwell equation. It should be noted that the NLS equation is parity-time-symmetry (PT-symmetry), which has becomed an interesting topic in quantum mechanics \cite{i1}, optics \cite{i4,i5}, Bose-Einstein condensates \cite{i6} and quantum chromodynamics \cite{i7}, etc.\par
A nonlocal NLS equation has been introduced by Ablowitz and Musslimani in \cite{i14}:
\begin{equation}\label{a2}
iq_t(x,t)=q_{xx}(x,t)\pm2q(x,t)q^\ast(-x,t)q(x,t).
\end{equation}
It can be yielded from the famous AKNS system. As the NLS equation \eqref{a1}, the nonlocal NLS equation \eqref{a2} is also PT-symmetric. It is an integrable system with the Lax pair. Ablowitz and Musslimani gave its infinitely many conservation laws and solved it through the inverse scattering transformation \cite{i14}. Eq.\eqref{a2} has different properties from eq.\eqref{a1}, e.g., eq.\eqref{a2} contains both bright and dark soliton \cite{i16} and solutions with periodic singularities \cite{i14}.\par
Very recently, motivated by the work of nonlocal NLS equation due to Ablowitz and Musslimani, we proposed and investigated a nonlocal modified Korteweg-de Vries (mKdV) equation in \cite{i28},
\begin{equation}\label{a4}
q_t(x,t)+6q(x,t)q(-x,-t)q_x(x,t)+q_{xxx}(x,t)=0.
\end{equation}
Its Lax integrability, Darboux transformation, and soliton solution have been discussed in our paper \cite{i28}. We should remark here that the nonlocal mKdV equation \eqref{a4} also occurred in a paper of Ablowitz and Musslimani \cite{i15}. It is obvious that the nonlocal mKdV equation \eqref{a4} with the reduction $q(-x,-t)=q(x,t)$ reduces to the mKdV equation.
%The modified Korteweg-de Vries equation (mKdV), that is
%\begin{equation}\label{a3}
%q_t(x,t)+6q^2(x,t)q_x(x,t)+q_{xxx}(x,t)=0,
%\end{equation}
The mKdV equation can be derived from Euler equation and has applications in varies physical fields \cite{i17,i18}. Wadati used inverse scattering transformation to study mKdV equation and obtained explicit solutions, including $N$-solitons, multiple-pole solutions and solutions derived from PT-symmetric potentials \cite{i23,i24,i25}. Hirota also achieved $N$-solitons by bilinear technique and investigated multiple collisions of solitons \cite{i26}.\par
In this paper, we will investigate further the new integrable nonlocal mKdV equation \eqref{a4}. We will construct exact solutions of the nonlocal mKdV equation \eqref{a4} including soliton and breather through inverse scattering transformation. These solutions have some new properties, which are different from the ones of the mKdV equation.\\
\section{Inverse scattering transformation on nonlocal mKdV equation}
The invention of inverse scattering transformation (IST) is due to the pioneering work of Gardner, Greene, Kruskal, and Miura for the Cauchy problem of KdV equation \cite{i21}. IST has been developed into a systematic method to achieve exact solutions for integrable nonlinear systems \cite{i29, i22,i27}. In this section, we will give the IST for the nonlocal mKdV equation \eqref{a4}. Start with the following linear problem,
\begin{align}
\varphi_x & =\textbf{U}\varphi=(-ik\sigma_3+\textbf{Q})\varphi,\label{1}\\
\varphi_t & =\textbf{V}\varphi=(-4ik^3\sigma_3+4k^2\textbf{Q}-2ik\textbf{V}_1+\textbf{V}_2)\varphi,\label{2}
\end{align}
with
\begin{gather*}
\sigma_3=\left(\begin{array}{cc} 1 & 0\\ 0 & -1\end{array}\right),\qquad
\textbf{Q}=\left(\begin{array}{cc} 0 & q(x,t)\\ r(x,t) & 0\end{array}\right),\\
\textbf{V}_1=(\textbf{Q}^2+\textbf{Q}_x)\sigma_3,\qquad\textbf{V}_2=-\textbf{Q}_{xx}+2\textbf{Q}^3+\textbf{Q}_x\textbf{Q}-\textbf{Q}\textbf{Q}_x,
\end{gather*}
where $\varphi=(\varphi_1(x,t),\varphi_2(x,t))^\textrm{T}$, and $k$ is the spectral parameter. The compatibility condition of system \eqref{1} and \eqref{2} $\textbf{U}_t-\textbf{V}_x+[\textbf{U},\textbf{V}]=0$ leads to
\begin{equation}\label{3}\begin{aligned}
& q_t(x,t)+q_{xxx}(x,t)-6q(x,t)r(x,t)q_x(x,t)=0,\\ & r_t(x,t)+r_{xxx}(x,t)-6q(x,t)r(x,t)r_x(x,t)=0.
\end{aligned}\end{equation}
Nonlocal mKdV equation \eqref{a4} is obtained from system \eqref{3} under the reduction
\begin{equation}\label{4}
r(x,t)=-q(-x,-t).
\end{equation}
Next, following the standard procedure of inverse scattering transformation(e.g. see [23],[24], [14]), we will give the inverse scattering for nonlocal mKdV equation. Assume $q(x,t)$ and its derivatives with respect to $x$ vanish rapidly at infinity. So does $r(x,t)$. Fix time $t=0$. Define $\phi(x,k)$ and $\bar{\phi}(x,k)$ as a pair of eigenfunctions of eq.\eqref{1}, which satisfy the following boundary conditions,
\begin{equation}\label{5}
\phi(x,k)\sim\left(\begin{array}{c}1\\0\end{array}\right)e^{-ikx},\quad\bar{\phi}(x,k)\sim\left(\begin{array}{c}0\\1\end{array}\right)e^{ikx},\quad
x\rightarrow-\infty.
\end{equation}
Similarly, $\psi(x,k)$ and $\bar{\psi}(x,k)$ are defined as another pair of eigenfunctions of eq.\eqref{1} satisfying a different boundary conditions,
\begin{equation}\label{6}
\psi(x,k)\sim\left(\begin{array}{c}0\\1\end{array}\right)e^{ikx},\quad\bar{\psi}(x,k)\sim\left(\begin{array}{c}1\\0\end{array}\right)e^{-ikx},\quad
x\rightarrow+\infty.
\end{equation}
Note that, in this paper, we denote the complex conjugation of $\phi$ by $\phi^\ast$ instead of $\bar{\phi}$. Furthermore, $\phi$ and $\psi$ are required to be analytic in upper half $k$-plane, while $\bar{\phi}$ and $\bar{\psi}$ are required to be analytic in lower half $k$-plane. For a solution $u(x,k)$ and $v(x,k)$ to eq.\eqref{1}, their Wronskian $W[u,v]=u_1v_2-u_2v_1$ is independent of $x$. Since $\{\phi,\bar{\phi}\}$ and $\{\psi,\bar{\psi}\}$ are linearly dependent, we set
\begin{equation}\label{7}\begin{aligned}
& \phi(x,k)=a(k)\bar{\psi}(x,k)+b(k)\psi(x,k),\\
& \bar{\phi}(x,k)=\bar{a}(k)\psi(x,k)+\bar{b}(k)\bar{\psi}(x,k).
\end{aligned}\end{equation}
The scattering data therefore can be expressed as
\begin{equation}\label{8}\begin{aligned}
& a(k)=W[\phi(x,k),\psi(x,k)], & b(k)=W[\psi(x,k),\bar{\phi}(x,k)],\\
& \bar{a}(k)=W[\bar{\psi}(x,k),\bar{\phi}(x,k)], & \bar{b}(k)=W[\bar{\phi}(x,k),\psi(x,k)].
\end{aligned}\end{equation}
One can prove that $\phi e^{ikx}$, $\psi e^{-ikx}$ and $a(k)$ are analytic functions in upper half $k$-plane; $\bar{\phi}e^{-ikx}$, $\bar{\psi}e^{ikx}$ and $\bar{a}(k)$ are analytic functions in lower half $k$-plane [23]. Define $\rho(k)=b(k)/a(k)$ and $\bar{\rho}(k)=\bar{b}(k)/\bar{a}(k)$ as reflection coefficients. Assume $k_m~(m=1,2,\cdots,N)$, the zeros of $a(k)$ in upper half $k$-plane, are single, as well as $\bar{k}_n~(n=1,2,\cdots,\bar{N})$ denoted as the zeros of $\bar{a}(k)$ in lower half $k$-plane. When $a(k_m)=0$, by eq.\eqref{8}, it yields that $\phi(x,k_m)$ and $\psi(x,k_m)$ are linearly dependent, i.e. there exist constants $\gamma_j$ such that $\phi(x,k_m)=\gamma_m\psi(x,k_m)$. Similarly, one has $\bar{\phi}(x,\bar{k}_n)=\bar{\gamma}_n\bar{\psi}(x,\bar{k}_n)$. The normalizing coefficients $\{c_m,\bar{c}_n\}$ are defined by
\begin{equation}\label{8a}
c_m^2=\dfrac{i\gamma_m}{\dot{a}(k_m)},~~(m=1,2,\cdots,N);\qquad\bar{c}_n^2=\dfrac{i\bar{\gamma}_n}{\dot{\bar{a}}(\bar{k}_n)},~~(n=1,2,\cdots,\bar{N}).
\end{equation}
We should note that, under the reduction \eqref{4}, the scattering data obeys $b(k)=-\bar{b}(-k^\ast)$, $a(k)=a^\ast(-k^\ast)$ and $\bar{a}(k)=\bar{a}^\ast(-k^\ast)$, when $q(x)$ is a real function. This means the eigenvalues are purely imaginary or appear in pairs $\{k_m, -k_m^\ast\}$ and $\{\bar{k}_n, -\bar{k}_n^\ast\}$.\par
Suppose the eigenfunctions $\psi$ and $\bar{\psi}$ satisfy the following forms:
\begin{equation}\label{9}\begin{aligned}
& \psi(x,k)=\left(\begin{array}{c}0\\1\end{array}\right)e^{ikx}+\int_x^{\infty}K(x,s)e^{iks}\textrm{d}s,\\
& \bar{\psi}(x,k)=\left(\begin{array}{c}1\\0\end{array}\right)e^{-ikx}+\int_x^{\infty}\bar{K}(x,s)e^{-iks}\textrm{d}s,
\end{aligned}\end{equation}
where $K(x,s)=(K_1(x,s),K_2(x,s))^\textrm{T}$ and $\bar{K}(x,s)=(\bar{K}_1(x,s),\bar{K}_2(x,s))^\textrm{T}$, $x<s$. Substituting eq.\eqref{9} into eq.\eqref{1} yields that $K_1(x,s)$ and $K_2(x,s)$ satisfy a Goursat problem, which means that the solution exists and is unique. Moreover, one can get the relations between potentials and $K(x,y)$ and $\bar{K}(x,y)$:
\begin{equation}\label{10}
q(x)=-2K_1(x,x),\qquad r(x)=-2\bar{K}_2(x,x).
\end{equation}
Let
\begin{equation}\label{11}\begin{array}{lll}
F_c(x)=\dfrac{1}{2\pi}\int_{-\infty}^{+\infty}\rho(k)e^{ikx}\textrm{d}k,
& F_d(x)=\underset{m=1}{\overset{N}{\sum}}c_m^2e^{ik_mx}, & F(x)=F_c(x)-F_d(x),\\
\bar{F}_c(x)=\dfrac{1}{2\pi}\int_{-\infty}^{+\infty}\bar{\rho}(k)e^{-ikx}\textrm{d}k,
& \bar{F}_d(x)=\underset{n=1}{\overset{\bar{N}}{\sum}}\bar{c}_n^2e^{-i\bar{k}_nx}, & \bar{F}(x)=\bar{F}_c(x)-\bar{F}_d(x).
\end{array}\end{equation}
Through eq.\eqref{7}, one achieves Gel'fand-Levitan-Marchenko integral equation (GLM):
\begin{equation}\label{12}\begin{aligned}
& \bar{K}(x,y)+\left(\begin{array}{c}0\\1\end{array}\right)F(x+y)+\int_x^{\infty}K(x,s)F(s+y)\textrm{d}s=0,\\
& K(x,y)+\left(\begin{array}{c}1\\0\end{array}\right)\bar{F}(x+y)+\int_x^{\infty}\bar{K}(x,s)\bar{F}(s+y)\textrm{d}s=0.
\end{aligned}\end{equation}
\indent The time evolution of scattering data $\{\rho(k,t),\bar{\rho}(k,t)\}$ and normalizing coefficients $\{c_m^2,\bar{c}_n^2\}$ are given by
\begin{equation}\label{13}\begin{array}{ll}
\rho(k,t)=\rho(k,0)e^{8ik^3t}, & c_m^2(t)=c_m^2(0)e^{8ik_m^3t}~~~(m=1,2,\cdots,N),\\
\bar{\rho}(k,t)=\bar{\rho}(k,0)e^{-8i\bar{k}^3t}, & \bar{c}_n^2(t)=\bar{c}_n^2(0)e^{-8i\bar{k}_n^3t}~~~(n=1,2,\cdots,\bar{N}).
\end{array}\end{equation}
Then, putting eq.\eqref{13} into eq.\eqref{11} and solving GLM eq.\eqref{12} yields $K(x,y;t)$ and $\bar{K}(x,y;t)$. Finally the solutions $q(x,t)$ and $r(x,t)$ are constructed. Assume the scattering problem is reflectionless, i.e. $\rho(k,t)=\bar{\rho}(k,t)\equiv 0$ and $K_1(x,y)$, $\bar{K}_2(x,y)$ have the following expressions:
\begin{equation}\label{14}
K_1(x,y)=\underset{n=1}{\overset{\bar{N}}{\sum}}\bar{c}_n(t)\bar{g}_n(x,t)e^{-i\bar{k}_ny},\qquad
\bar{K}_2(x,y)=\underset{m=1}{\overset{N}{\sum}}c_m(t)g_m(x,t)e^{ik_my}.
\end{equation}
Introduce $N\times 1$ column vector $h(x,t)=(h_1(x,t),...h_m(x,t),..., h_N(x,t))^T$, $\bar{N}\times 1$ column vector $\bar{h}(x,t)=(\bar{h}_1(x,t),...\bar{h}_n(x,t),..., \bar{h}_{\bar{N}}(x,t))^T$ and matrix $E(x,t)=(e_{nm})_{\bar{N}\times N}$, where
\begin{equation*}
h_m(x,t)=c_m(t)e^{ik_mx},\qquad \bar{h}_n(x,t)=\bar{c}_n(t)e^{-i\bar{k}_nx},\qquad e_{nm}(x,t)=\dfrac{h_m(x,t)\bar{h}_n(x,t)}{k_m-\bar{k}_n}.
\end{equation*}
After some calculations, $q(x,t)$ and $r(x,t)$ are written
\begin{equation}\label{15}\begin{aligned}
& q(x,t)=-2\textrm{tr}\left((I_{\bar{N}}+E(x,t)E(x,t)^{\textrm{T}})^{-1}\bar{h}(x,t)\bar{h}(x,t)^{\textrm{T}}\right),\\
& r(x,t)=-2\textrm{tr}\left((I_N+E(x,t)^{\textrm{T}}E(x,t))^{-1}h(x,t)h(x,t)^{\textrm{T}}\right),
\end{aligned}\end{equation}
where $I_N$ or $I_{\bar{N}}$ is a $N$-dimensional or $\bar{N}$-dimensional unit matrix. When eigenvalues $\{k_m,\bar{k}_n\}$ are suitably selected and eq.\eqref{15} satisfies the constraint \eqref{4}, $q(x,t)$ becomes the solution of eq.\eqref{a4} with initial scattering data $\{c_m(0),\bar{c}_n(0)\}$.\\
We will emphasize here that the procedure described above of solving nonlocal mKdV equation seems same as the one for the classical mKdV equation, but there exists important difference between these two cases.  The scattering coefficients $a(k)$ and $\bar{a}(k)$ for the nonlocal case have no relations, while ones of classical problems have. This leads to that eigenvalues $k_j, \bar{k}_j$ are not related, either. The normalizing coefficients $c_j, \bar{c}_j$ depend on the eigenvalues $k_j, \bar{k}_j$ in the nonlocal case, which will be mentioned in the next section, rather than being free parameters in the classical case. In the classical case, eigenfunctions, which are analytic in the upper $k$-plane, are related to those being analytic in the lower $k$-plane. But, this property does not hold anymore in the nonlocal case. This is the most important difference between these two cases, which is also mentioned in \cite{i15}.
\section{Soliton solutions and their properties}
In this section, we will derive soliton solutions of integrable nonlocal mKdV equation (3) from the explicit formula \eqref{15}.\\
{\bf Case 1. one-soliton solutions}\\
Let $N=\bar{N}=1$ and the eigenvalues be purely imaginary. From formula \eqref{15} and the symmetry reduction \eqref{4}, it can be derived that $c_1(0), \bar{c}_1(0)$ and $k_1, \bar{k}_1$ have the following constraints:
\begin{equation}
(k_1-\bar{k}_1)^2+(c_1(0))^4=0,\qquad(k_1-\bar{k}_1)^2+(\bar{c}_1(0))^4=0.
\end{equation}
Denote $k=i\alpha$ and $\bar{k}=-i\beta$, where $\alpha,\beta>0$. Substituting the above constraints into eq.\eqref{15} yields the one-soliton solution
\begin{equation}\label{16}
q(x,t)=\dfrac{2(\alpha+\beta)}{e^{-2\alpha(x-4\alpha^2t)}+\sigma e^{2\beta(x-4\beta^2t)}},
\end{equation}
where $\sigma=\pm 1$. Let $\sigma=1$, $q$ can be written
\begin{equation}\label{17}
q(x,t)=(\alpha+\beta)e^{(\alpha-\beta)x-4(\alpha^3-\beta^3)t}~\textrm{sech}((\alpha+\beta)x-4(\alpha^3+\beta^3)t).
\end{equation}
It is obvious that for arbitrary fixed $t$, $q(x,t)\rightarrow 0$ as $|x|\rightarrow \infty$. This solution $q(x,t)$ of nonlocal mKdV equation is a soliton solution, but it has different property from the one of classical mKdV equation. We note that, when $x$ and $t$ satisfy $x/t=k+o(t^{-1})~(t\rightarrow\infty)$, where $k$ is a constant between $4\alpha^2$ and $4\beta^2$, $q(x,t)$ goes to infinity along these directions as $t\rightarrow+\infty$ for $\alpha<\beta$, or $t\rightarrow-\infty$ for $\alpha>\beta$. It indicates that $q(x,t)$ evolves like a solitary wave with its amplitude increasing or decaying exponentially. Fig. 1 describes this property. We can see that $q(x,t)$ is a usual soliton in the case of $\alpha=\beta$. Notice that in this case $q(x,t)=q(-x,-t)$, and $\bar{k}_1=k_1^*$. This means that $q(x,t)$ is also a soliton solution to mKdV equation. If $\sigma=-1$,
\begin{equation}\label{17a}
q(x,t)=-(\alpha+\beta)e^{(\alpha-\beta)x-4(\alpha^3-\beta^3)t}~\textrm{csch}((\alpha+\beta)x-4(\alpha^3+\beta^3)t).
\end{equation}
So, $q(x,t)$ possesses singularity at the line $\{(x,t)|x=4(\alpha^2-\alpha\beta+\beta^2)t\}$.\\
{\bf Case 2. two-soliton solutions}\\
Set $N=\bar{N}=2$. First, we obtain the constraints between normalizing coefficients and eigenvalues via eq.\eqref{15} and eq.\eqref{4} by direct calculations:
\begin{equation}\begin{aligned}
& (c_1(0))^4+\dfrac{(\bar{k}_1-k_1)^2(\bar{k}_2-k_1)^2}{(k_1-k_2)^2}=0,\quad(c_2(0))^4+\dfrac{(\bar{k}_1-k_2)^2(\bar{k}_2-k_2)^2}{(k_1-k_2)^2}=0,\\
& (\bar{c}_1(0))^4+\dfrac{(k_1-\bar{k}_1)^2(k_2-\bar{k}_1)^2}{(\bar{k}_1-\bar{k}_2)^2}=0,\quad
(\bar{c}_2(0))^4+\dfrac{(k_1-\bar{k}_2)^2(k_2-\bar{k}_2)^2}{(\bar{k}_1-\bar{k}_2)^2}=0.
\end{aligned}\end{equation}
Then, the general expression of a two-soliton solution is
\begin{equation}\label{18}\begin{aligned}
q(x,t) & =-2i\dfrac{F(x,t)}{G(x,t)},\\
F(x,t) & =\dfrac{\bar{\sigma}_1(k_1-\bar{k}_1)(k_2-\bar{k}_1)}{\bar{k}_1-\bar{k}_2}e^{\bar{\xi}_1}
+\dfrac{\bar{\sigma}_2(k_1-\bar{k}_2)(k_2-\bar{k}_2)}{\bar{k}_1-\bar{k}_2}e^{\bar{\xi}_2}\\
&\quad -\dfrac{\sigma_1\bar{\sigma}_1\bar{\sigma}_2(\bar{k}_1-k_2)(\bar{k}_2-k_2)}{k_1-k_2}e^{\xi_1+\bar{\xi}_1+\bar{\xi}_2}
-\dfrac{\sigma_2\bar{\sigma}_1\bar{\sigma}_2(\bar{k}_1-k_1)(\bar{k}_2-k_1)}{k_1-k_2}e^{\xi_2+\bar{\xi}_1+\bar{\xi}_2},\\
G(x,t) & =1-\dfrac{(k_1-\bar{k}_2)(k_2-\bar{k}_1)}{(k_1-k_2)(\bar{k}_1-\bar{k}_2)}\left(\sigma_1\bar{\sigma}_1e^{\xi_1+\bar{\xi}_1}
+\sigma_2\bar{\sigma}_2e^{\xi_2+\bar{\xi}_2}\right)\\
&\quad -\dfrac{(k_1-\bar{k}_1)(k_2-\bar{k}_2)}{(k_1-k_2)(\bar{k}_1-\bar{k}_2)}\left(\sigma_1\bar{\sigma}_2e^{\xi_1+\bar{\xi}_2}
+\sigma_2\bar{\sigma}_1e^{\xi_2+\bar{\xi}_1}\right)+\sigma_1\sigma_2\bar{\sigma}_1\bar{\sigma}_2e^{\xi_1+\xi_2+\bar{\xi}_1+\bar{\xi}_2},
\end{aligned}\end{equation}
where $\sigma_j,\bar{\sigma}_j=\pm 1~(j=1,2)$ and
\begin{equation*}
\xi_j=2ik_j(x+4k_j^2t),\qquad \bar{\xi}_j=-2i\bar{k}_j(x+4\bar{k}_j^2t),~~(j=1,2).
\end{equation*}
Here, we focus on the case of $\{k_j,\bar{k}_j\}_{j=1}^2$ being purely imaginary. Set $k_j=i\alpha_j$, $\bar{k}_j=-i\beta_j$, where $\alpha_j,\beta_j>0$, and $\sigma_j=\bar{\sigma}_j=1~(j=1,2)$. For $(\alpha_1-\alpha_2)(\beta_1-\beta_2)>0$, eq.\eqref{18} is simplified to
\begin{equation}\label{19}\begin{aligned}
q(x,t) & =\dfrac{2F_1(x,t)}{G_1(x,t)},\\
F_1(x,t) & =A[(\alpha_1+\beta_1)e^{u_{2-}}\cosh(u_{2+}+\theta_2)+(\alpha_2+\beta_2)e^{u_{1-}}\cosh(u_{1+}+\theta_1)],\\
G_1(x,t) & =e^{u_{1-}+u_{2-}}[(\alpha_2-\alpha_1)(\beta_2-\beta_1)\cosh(u_{1+}+u_{2+})\\
& \quad+(\alpha_1+\beta_1)(\alpha_2+\beta_2)\cosh(u_{1-}-u_{2-})+(\alpha_1+\beta_2)(\alpha_2+\beta_1)\cosh(u_{1+}-u_{2+})],
\end{aligned}\end{equation}
where
\begin{gather*}
u_{j\pm}=\dfrac{1}{2}(\xi_j\pm\bar{\xi}_j),\qquad A=\sqrt{(\alpha_2-\alpha_1)(\beta_2-\beta_1)(\alpha_1+\beta_2)(\alpha_2+\beta_1)},\\
e^{\theta_1}=\dfrac{A}{|\alpha_2-\alpha_1|(\alpha_2+\beta_1)},\qquad e^{\theta_2}=\dfrac{A}{|\alpha_2-\alpha_1|(\alpha_1+\beta_2)}.
\end{gather*}
This is a two-soliton solution. In fig. 2, we describe such a two-soliton with $\alpha_1<\beta_1$ and $\alpha_2=\beta_2$. In this case, we see that the amplitude of one solitary wave has exponential increase as $t\rightarrow+\infty$, and another amplitude is stable but has a change during the collision of the two solitary waves. Furthermore, after interaction of the two solitary waves, there is a shift of phase and no change in the speed of them. Fig. 3 gives the case of $\alpha_1<\beta_1$ and $\alpha_2>\beta_2$, i.e., the amplitude of a solitary wave increases exponentially, and the one of another solitary wave decreases exponentially. The all solutions above belong to the interactions of bright-bright solitons. Interactions of bright-dark solitons can be found by setting $\sigma_2=-1$ and $\bar{\sigma}_2=-1$. The results are similar to the bright-bright case. In fig. 4,  we give an example of the increase-increase case, i.e., the amplitudes of both two solitary waves have exponential increase as $t\rightarrow+\infty$, and the amplitude below zero increases faster than the one above zero. During the interaction, both two solitary waves have a shift of the phase respectively and no changes in speed. In the case of $\alpha_j=\beta_j$, i.e., $\bar{k}_j=k_j^*$, $(j=1,2)$, the solution is a usual 2-soliton solution to nonlocal mKdV equation \eqref{a4} as well as to mKdV equation. For the case of $(\alpha_1-\alpha_2)(\beta_1-\beta_2)<0$, the solution always has singularity at some sites.\\
{\bf Case 3. Breather solution}\\
Let us consider the case of $k_1=-k_2^\ast$, $\bar{k}_1=-\bar{k}_2^\ast$, where $k_1$ and $\bar{k}_1$ are denoted by $k_1=\eta_1+i\zeta_1$ and $\bar{k}_1=\eta_2-i\zeta_2$ ( $\eta_j,\zeta_j, ~j=1,2$ are positive), and $\sigma_1\sigma_2=-1$ and $\bar{\sigma}_1\bar{\sigma}_2=-1$. In this case,
 the solution has the expression,
\begin{equation}\label{19a}\begin{aligned}
q(x,t) & =\dfrac{2F_2(x,t)}{G_2(x,t)},\\
F_2(x,t) & =\eta_1[(\eta_1^2-\eta_2^2+(\zeta_1+\zeta_2)^2)\sin v_{2+}-2\eta_2(\zeta_1+\zeta_2)\cos v_{2+}]e^{-v_{1-}}\\
& \quad+\eta_2[(\eta_1^2-\eta_2^2-(\zeta_1+\zeta_2)^2)\sin v_{1+}-2\eta_1(\zeta_1+\zeta_2)\cos v_{1+}]e^{v_{2-}},\\
G_2(x,t) & =2\eta_1\eta_2\cosh(v_{1-}+v_{2-})+2\eta_1\eta_2(1+\cos v_{1+}\cos v_{2+})\\
& \quad+[\eta_1^2+\eta_2^2+(\zeta_1+\zeta_2)^2]\sin v_{1+}\sin v_{2+},
\end{aligned}\end{equation}
where
$$
v_{j+}=2\eta_j[x+4(\eta_j^2-3\zeta_j^2)t],\quad v_{j-}=-2\zeta_j[x+4(3\eta_j^2-\zeta_j^2)t],\quad(j=1,2).
$$
The solution possesses singularity if $\eta_1\neq\eta_2$ or $\zeta_1\neq\zeta_2$. But, selecting $\eta_1=\eta_2=\zeta_1=\zeta_2$ in eq.\eqref{19a} yields an interesting solution,
\begin{equation}\label{20}
q(x,t)=4\mu\dfrac{\sinh(\xi_+)\sin(\xi_-)-\cosh(\xi_+)\cos(\xi_-)}{\cosh^2(\xi_+)+\sin^2(\xi_-)},
\end{equation}
where $\xi_\pm=-2\mu(x\pm 8\mu^2t)$ with $\mu>0$. This is a breather solution (see fig. 5).
\section{Conclusions and discussions}
In this paper, we have investigated the nonlocal mKdV equation through inverse scattering method. We have given its solutions in the general form. We have presented one-soliton, two-soliton and breather solutions. The analysis of the properties of these solutions has been given, including the singularity and long-time behavior. We have demonstrated that these solutions for nonlocal mKdV equation have some different properties from ones of mKdV equation. In Ref. \cite{i15}, Ablowitz and  Musslimani introduced the other two integrable nonlocal equations, complex nonlocal mKdV equation, and nonlocal sine-Gordon equation. We will give inverse scattering transformations and soliton solutions for the two new integrable nonlocal equations in the future work.

\begin{figure*}\label{p1}
\centering
\subfigure[]{\includegraphics[width=1.5in]{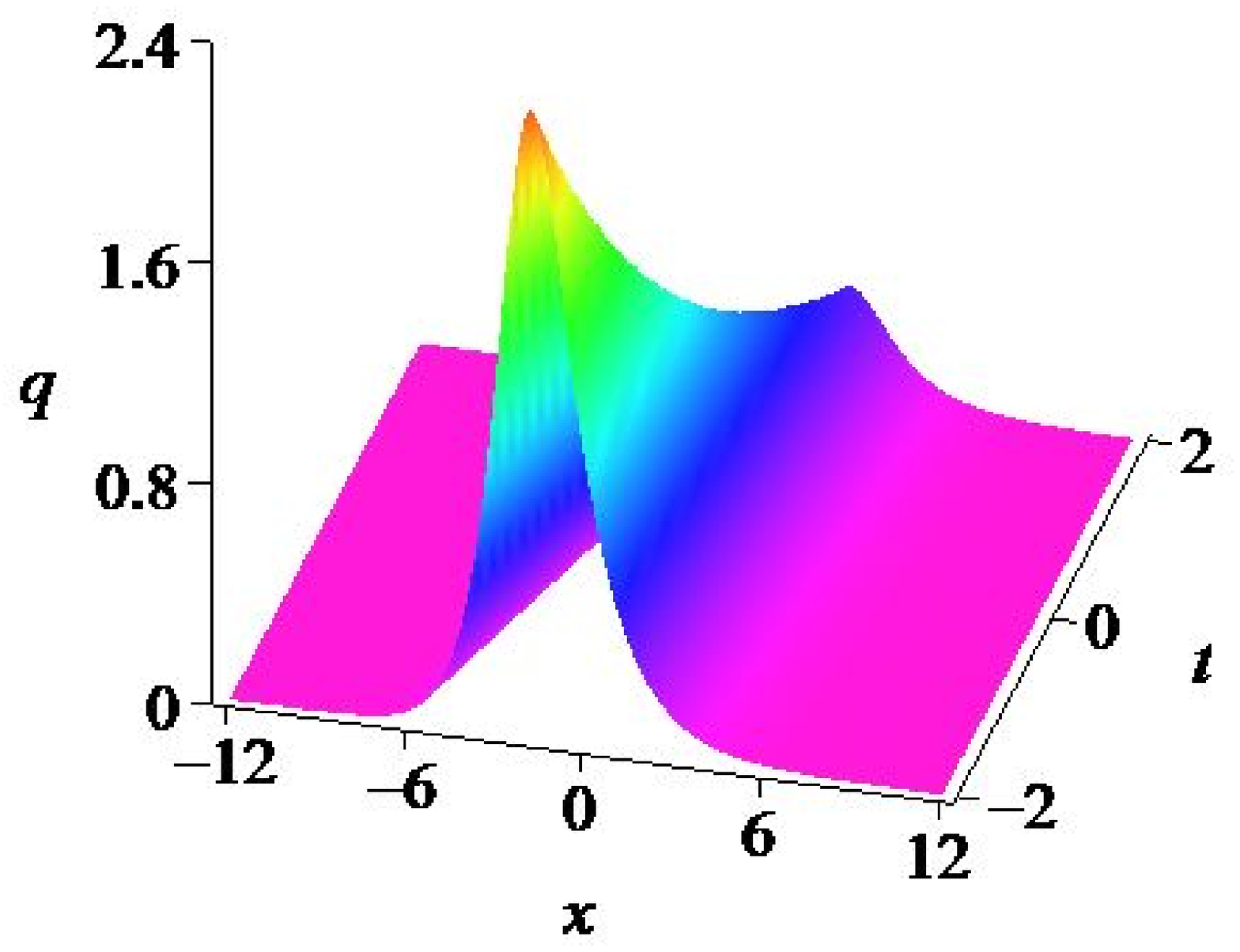}}\qquad
\subfigure[]{\includegraphics[width=1.5in]{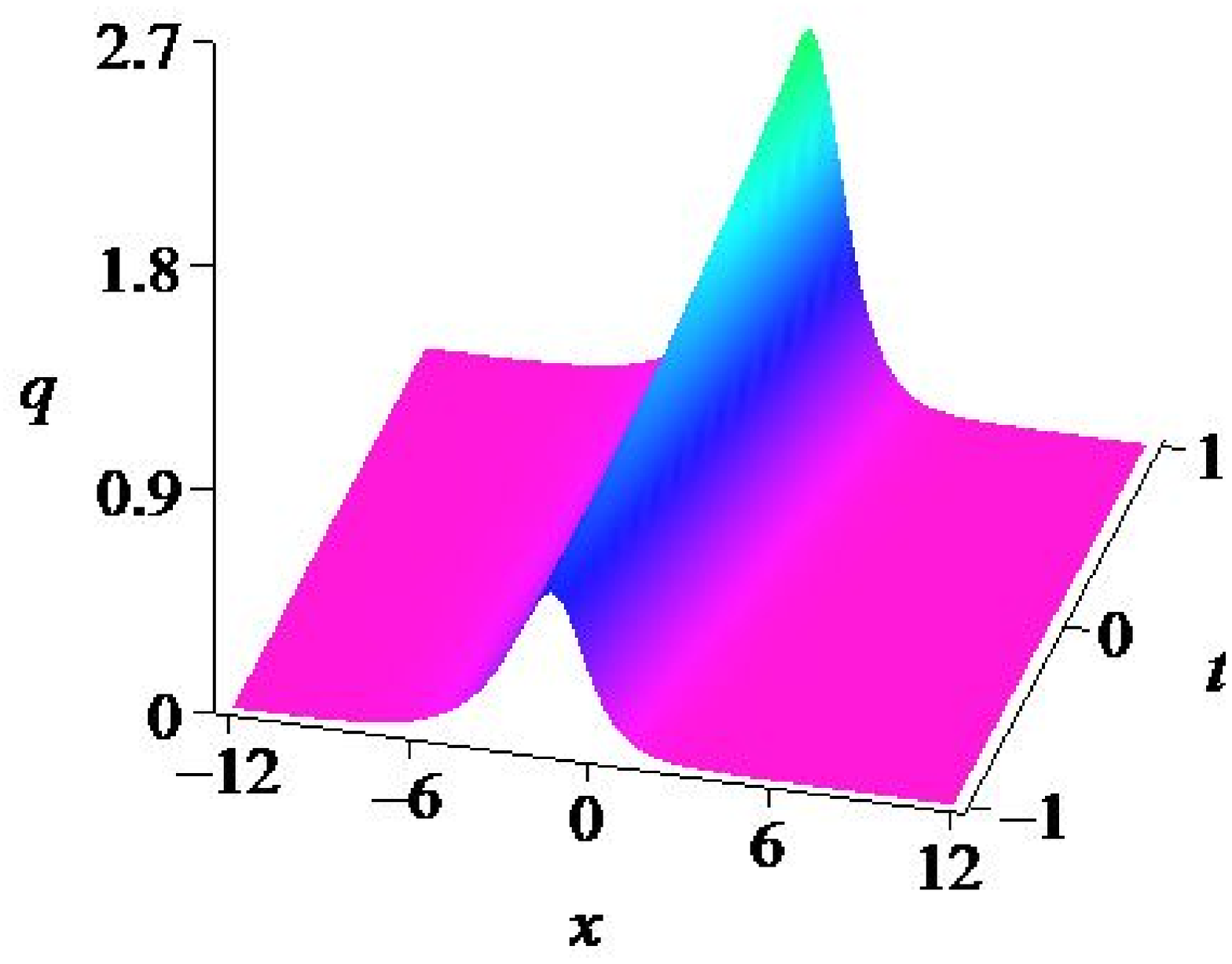}}
\caption{\small{(a) one-soliton-like solution given by eq.\eqref{17} with $\alpha=3/5$ and $\beta=1/3$. The amplitude decays exponentially as $t$ increases.;(b)
one-soliton-like solution given by eq.\eqref{17} with $\alpha=1/3$ and $\beta=3/5$. The amplitude increases exponentially as $t$ increases.}}
\end{figure*}

\begin{figure*}\label{p2}
\centering
\subfigure[]{\includegraphics[width=1.5in]{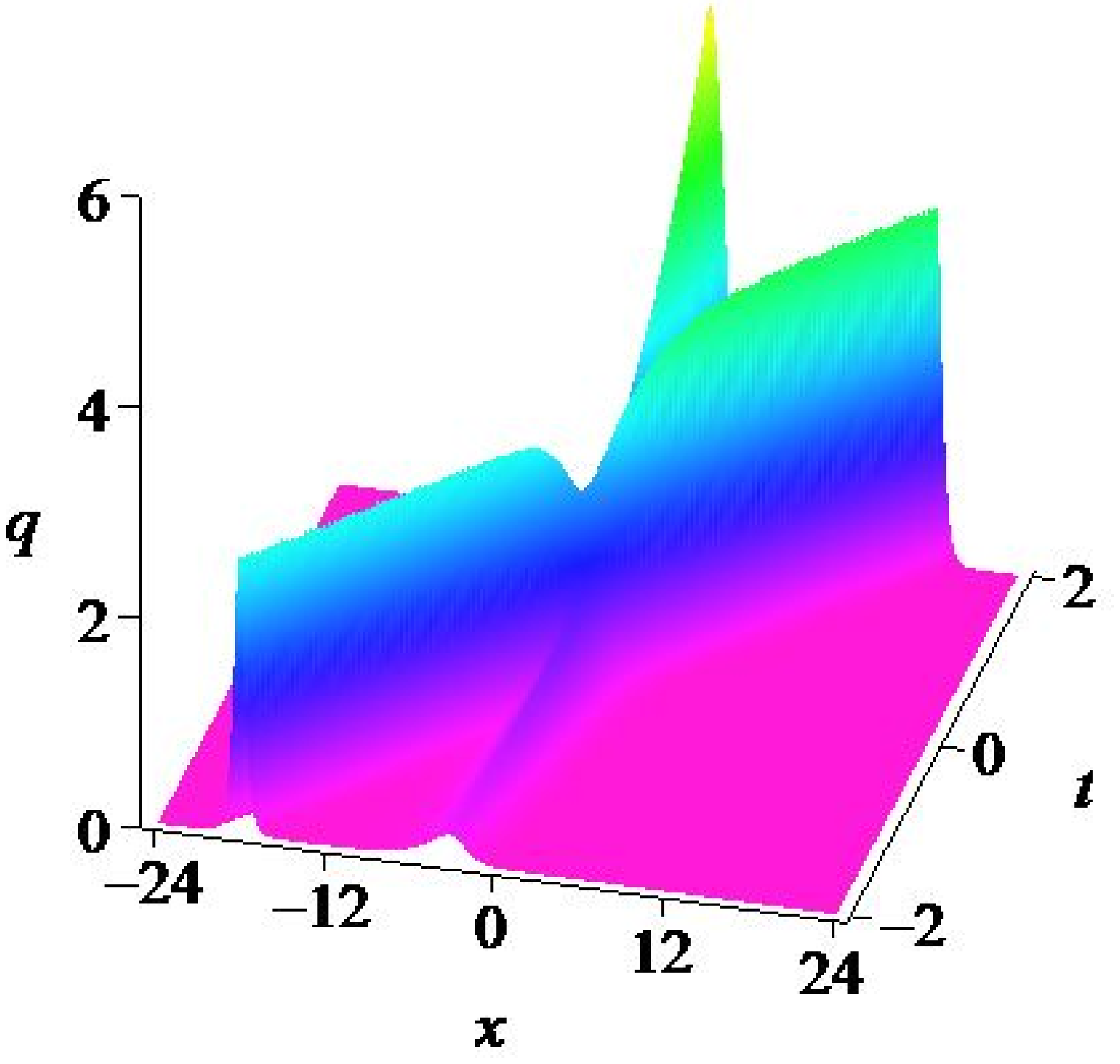}}\qquad
\subfigure[]{\includegraphics[width=1.0in]{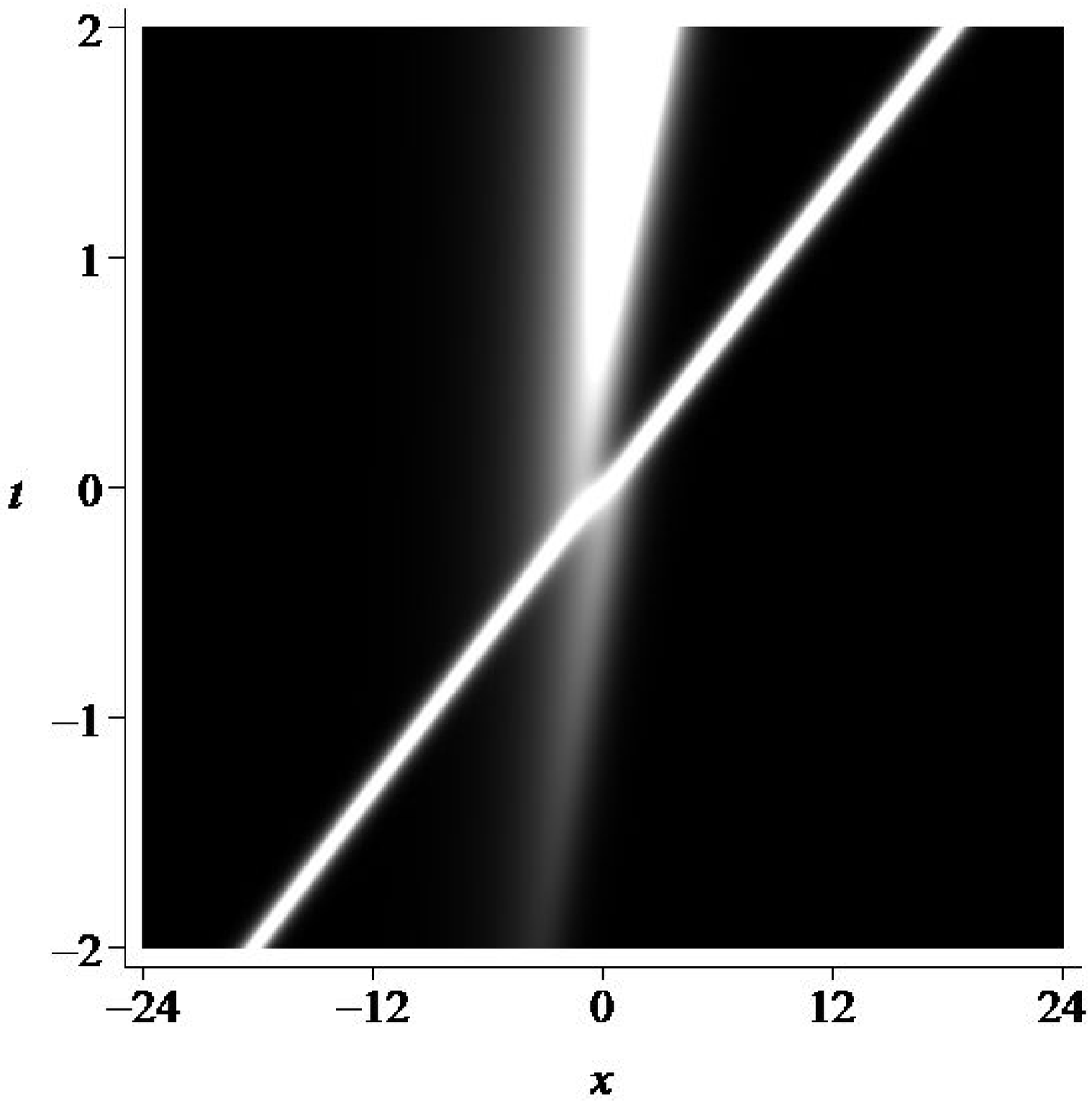}}
\caption{\small{two-soliton-like solution of bright-bright kind given by eq.\eqref{19} with $\alpha_1=1/4$, $\beta_1=3/4$ and $\alpha_2=\beta_2=3/2$. Only one of the amplitudes increases exponentially as $t$ increases.}}
\end{figure*}

\begin{figure*}\label{p3}
\centering
\subfigure[]{\includegraphics[width=1.5in]{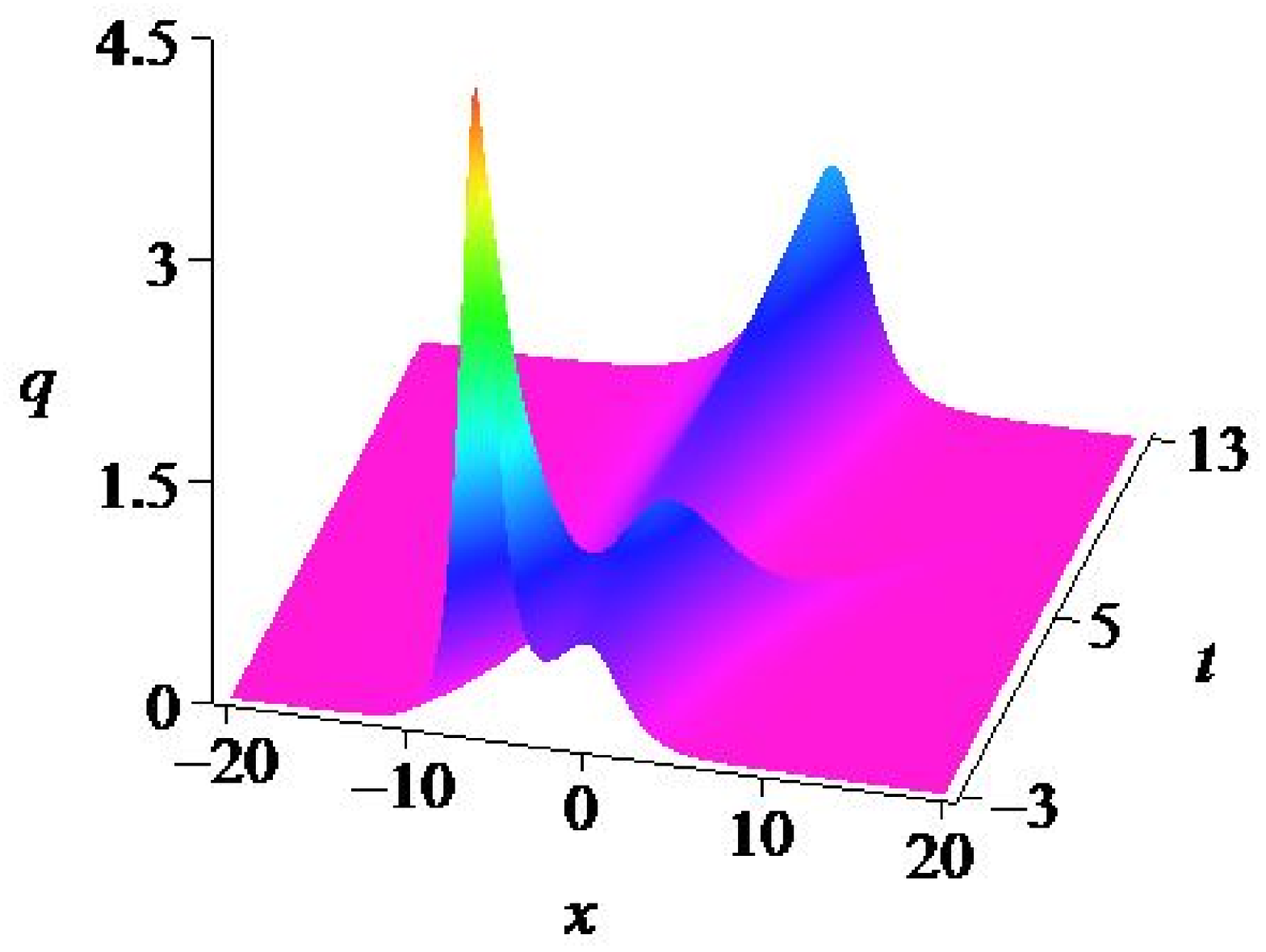}}\qquad
\subfigure[]{\includegraphics[width=1.0in]{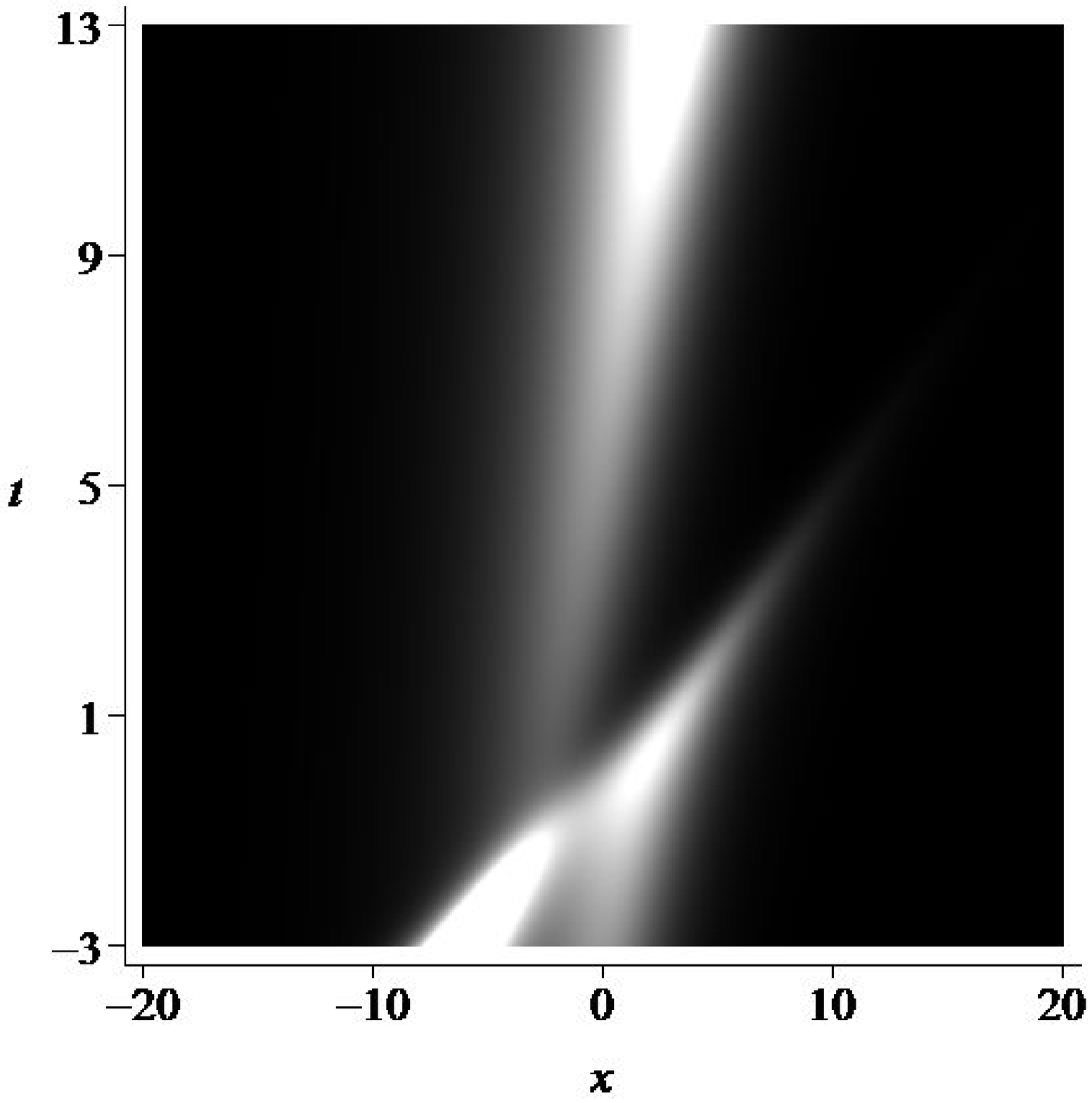}}
\caption{\small{two-soliton-like solution of bright-bright kind given by eq.\eqref{19} with $\alpha_1=3/16$, $\beta_1=3/8$, $\alpha_2=3/4$ and $\beta_2=9/16$. One of the amplitudes increases exponentially and the other decrease exponentially as $t$ increases.}}
\end{figure*}

\begin{figure*}\label{p5}
\centering
\subfigure[]{\includegraphics[width=1.5in]{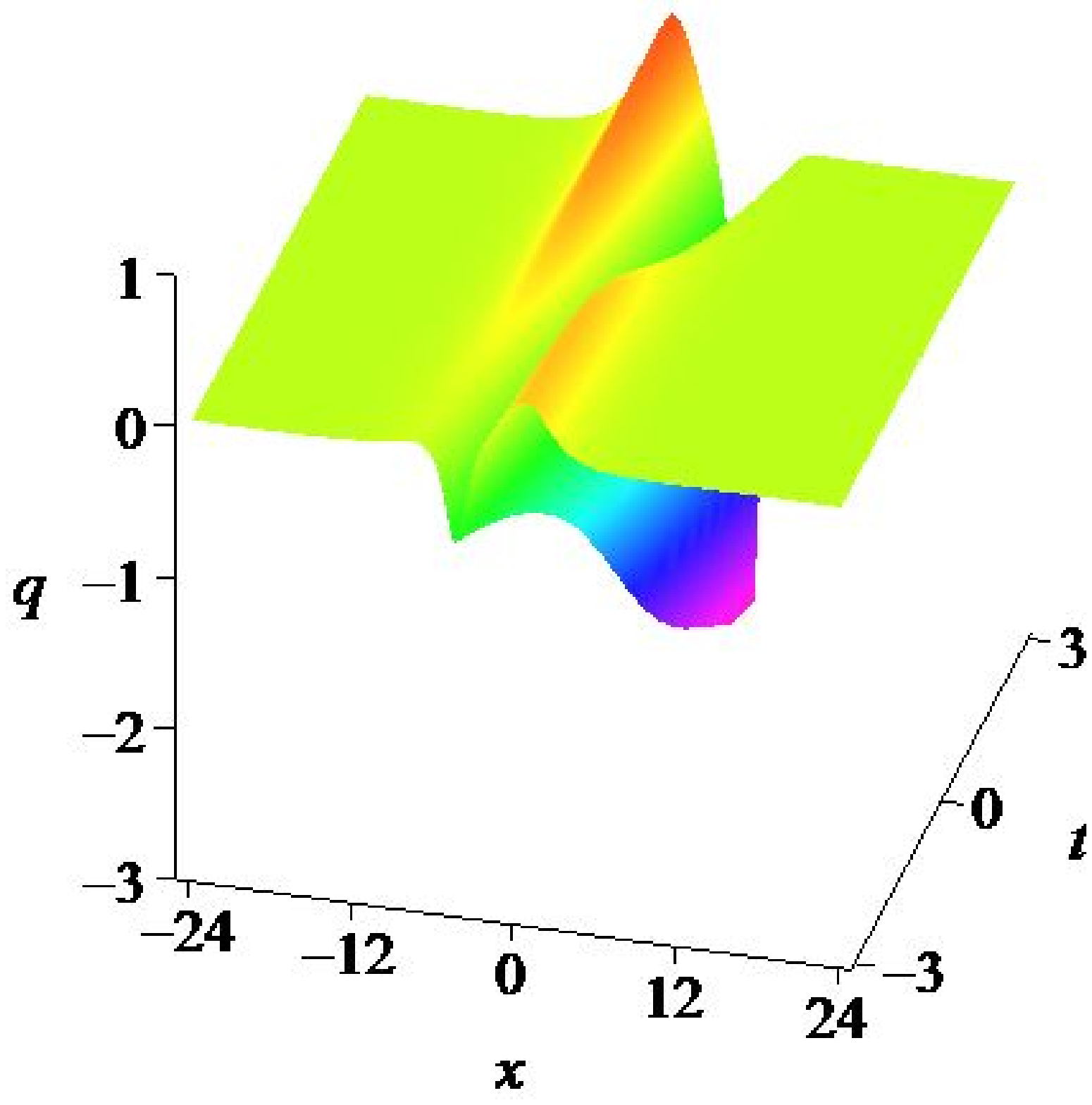}}\qquad
\subfigure[]{\includegraphics[width=1.0in]{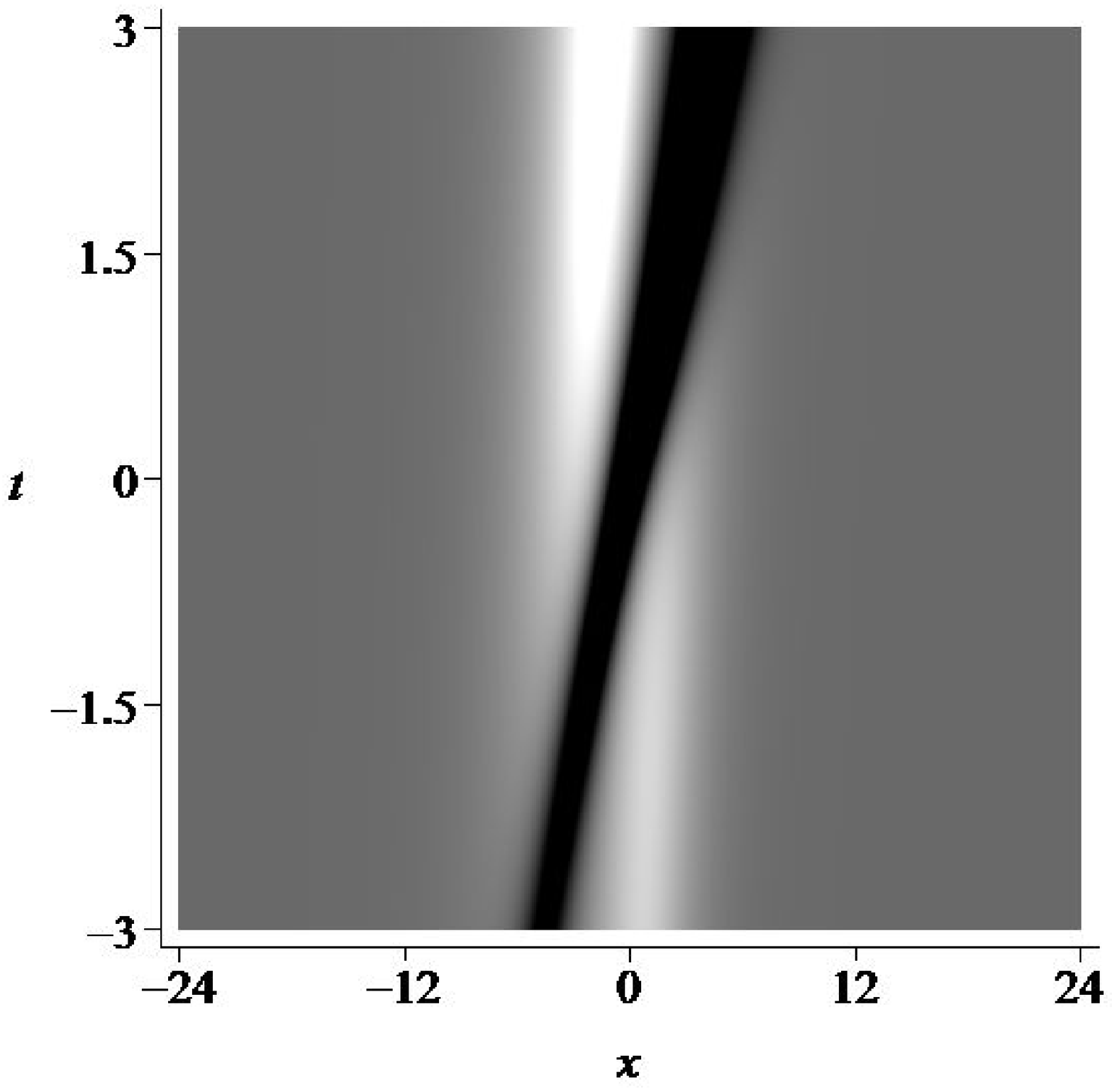}}
\caption{\small{two-soliton solution of bright-dark kind given by eq.\eqref{18} with $\sigma_1=1$, $\sigma_2=-1$, $\bar{\sigma}_1=1$ and $\bar{\sigma}_2=-1$, $k_1=i/2$, $\bar{k}_1=-i/3$, $k_2=i/4$ and $\bar{k}_2=-3i/5$.}}
\end{figure*}

\begin{figure*}\label{p6}
\centering
\subfigure[]{\includegraphics[width=1.5in]{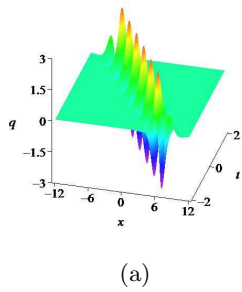}}
\caption{\small{Breather solution given by eq.\eqref{20} with $\mu=2/3$.}}
\end{figure*}

\vskip 16pt \noindent {\bf
Acknowledgements} \vskip 12pt
The work of ZNZ is supported by the National Natural Science
Foundation of China under grants 11271254 and 11428102, and
in part by the Ministry of Economy and Competitiveness of Spain under
contract MTM2012-37070.\\

\end{document}